\documentclass[10pt]{iopart}

\usepackage{iopams}
\usepackage{graphicx}

\begin{document}

\title[Experimental and computational characterization of a modified GEC cell.]{Experimental and computational characterization of a modified GEC cell for dusty plasma experiments.}

\author{Victor Land${}^1$, Erica Shen${}^{2}$, Bernard Smith${}^1$, \\Lorin Matthews${}^{1}$, and Truell Hyde${}^1$}

\address{$^1$ Center for Astrophysics, Space Physics, and Engineering Research, Baylor University, Waco, TX, USA 76798-7316\\
$^2$ Participant in Baylor University's High School Scholars Summer Research Program}
\ead{victor\_land@baylor.edu}

\begin{abstract}
A self-consistent fluid model developed for simulations of micro-gravity dusty plasma experiments has for the first time
been used to model asymmetric dusty plasma experiments in a modified GEC reference cell with gravity. The numerical results
are directly compared with experimental data and the experimentally determined dependence of global discharge
parameters on the applied driving potential and neutral gas pressure is found to be well matched by the model. The local profiles important for dust particle
transport are studied and compared with experimentally determined profiles. The radial forces in the midplane are
presented for the different discharge settings. The differences between the results obtained in the modified GEC cell and the results first reported for the original GEC reference cell are pointed out.
\end{abstract}


\maketitle
\section{Motivation for this study}

The GEC reference cell was originally designed to allow fair comparison between plasma processing studies performed in
different laboratories \cite{GEC1,GEC2}. A large experimental and numerical effort was undertaken to understand both the proper
operation of the cell from a technical viewpoint, as well as the physics of chemically active plasmas with different gas
mixtures \cite{GEC3,GEC4,GEC5}.

In the early nineties, it was realized that the GEC cell could also play the role of a standard experimental platform for
dusty plasma experiments \cite{GEC-dust1}. Several modifications to the original design were required in order to suspend dust particles in the discharge and to use different optical systems to visualize them. This in a way led to a loss of standardization, since different solutions were to deal with the additional challenge of conducting dusty plasma experiments \cite{GEC-dust2}.

Despite numerous experimental and numerical efforts to describe these experiments, a study employing a self-consistent numerical model,
which can self-consistently calculate the dust forces from the plasma parameters, is missing. Furthermore, the changes in discharge characteristics with respect to the original GEC reference cell, due to the necessary modifications for dusty plasma experiments, are usually ignored.

We have developed such a code in the past and applied it to micro-gravity dusty plasma experiments in
symmetrically driven discharges as well as to devices including the effects of gravity and additional thermophoretic forces, due to heated surfaces
\cite{model-1,model-2, model-3, model-4}. The
results have always shown excellent agreement with results reported in the literature, but the model has never been directly compared to measurements in a GEC reference cell.

The motivation for this study is therefore to examine dusty plasma environments in a modified GEC
cell with a self-consistent dusty plasma model for the first time, to compare results from the model to
measurements of plasma properties in the experiment, rather than from the literature alone, and determine the effect of the modifications to the GEC cell on the local and global discharge characteristics. The latter depend on the global particle and power balance of the discharge, which can be observed through the DC bias on the powered electrode and the power absorbed in the plasma. The local parameters include the dust charge, the plasma densities, and the plasma potential, which directly determine the forces that would act on dust particles present in the discharge.

Section \ref{sec:experiment} describes the geometry of the modified GEC
cell used at CASPER, and clarifies the Langmuir probe measurements, section \ref{sec:model} discusses the numerical model, section \ref{sec:globalresults} shows the results for the
global parameters studied, i.e. the DC bias and the absorbed power and section \ref{sec:localresults} shows the results for the local plasma profiles, i.e. the plasma potential, the particle densities, and the derived dust charge number. In section \ref{sec:forces} the forces
obtained from the measured plasma parameters are presented and compared to the outcome of the model. Section \ref{sec:summary} discusses the results and briefly mentions the outlook for future work.

\section{Description of the experimental setup}\label{sec:experiment}

\subsection{The modified GEC cell.}

\begin{figure}[htbp]
\begin{center}
\includegraphics[width=0.6\textwidth]{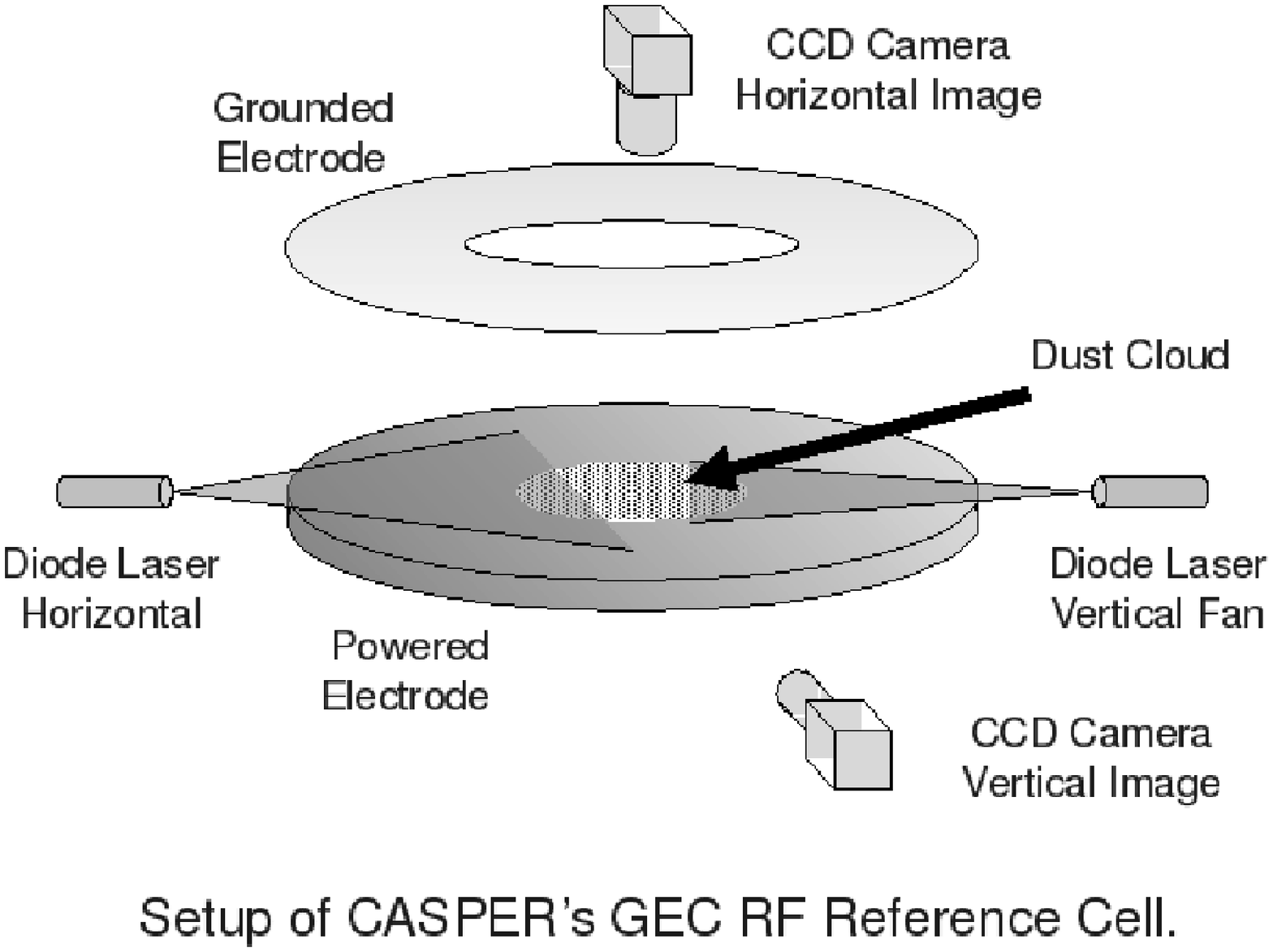}
\caption{Sketch of the interior of the modified GEC cell. The lower electrode is powered, the upper
electrode is grounded, as are the outer walls and the groundshield surrounding the lower electrode. The inter-electrode spacing is 3 cm, the radius of the cell is 13 cm, the height is 21 cm, and the electrode radius is 5.4 cm. The cutouts used for this experiment were 0.63, 1.25, or 2.5 cm in radius and 1 mm deep.}\label{fig:1}
\end{center}
\end{figure}

The GEC reference cell used by CASPER, shown schematically in figure \ref{fig:1}, is modified to allow dusty plasma experiments to be performed. The upper electrode is a grounded hollow cylinder rather than a solid electrode, so that a top-mounted camera can be used to take pictures of dust crystals from above. In total, two camera/dye laser systems have been added, to capture side-view and top-view pictures of dust clouds suspended in the plasma volume. The lasers are equipped with cylindrical lenses,  to creat thin laser sheets that illuminate selected areas within the dust clouds. The cameras can also be equipped with filters that allow only light at the laser frequency to pass through. This helps to select light scattered by the dust particles, and not from the plasma glow. 

The dust clouds suspended in the modified GEC cell are confined in the radial direction by a parabolically shaped electric potential, created by a circular cutout milled in a cover plate set on top of the powered electrode. The different cutouts used in the experiments have radii of 0.63, 1.25, and 2.5 cm. To introduce particles into the plasma, two dust shakers have been added to the top flange near the upper grounded electrode. Tapping these shakers forces dust particles to enter the plasma under the force of gravity. The bottom of each dust shakers is covered by a calibrated mesh to prevent larger clumps from entering the plasma.

The discharge parameters under external control include the neutral gas pressure, which can be adjusted using a butterfly valve controlling the
input gas flow, the input power, which is adjusted by changing the driving potential of the RF source, and the DC bias on
the powered electrode. For the current study, the DC bias was allowed to float.

\subsection{Langmuir probe.}

The plasma parameters in the modified GEC cell described above were measured with a SmartProbe produced by Scientific Instruments LTD \cite{SmartProbe}. The SmartProbe is a RF-compensated Langmuir probe, with a 10 mm long, 0.38 mm diameter tungsten probe tip attached to a 470 mm long shaft. By moving the shaft toward or away from the center of the powered lower electrode, radial profiles of the plasma parameters were obtained at the midplane between the two electrodes. The probe was inserted into the system through a sideport in an attempt to ensure that the profiles were obtained exactly in this plane; however, bending of the probeshaft by gravity can not be excluded. The deviation from this plane was estimated to be less than 10\% of the inter-electrode spacing, after inspection of side-view pictures. 

The radial probe position with respect to the electrode center was determined through examination of still-frame images from the top-mounted camera. Using imaging software \cite{ImageJ}, the midpoint of the 10 mm long probe tip and the center of the electrode were determined. The line connecting these two points was taken to be the true radial direction and any small angle the tip made with this line was measured. The projection of the probe tip onto this line was then taken as the error in the radial direction, assuming that the probe measurements represent the plasma parameters averaged over the length of the tip.

For each radial position, the bias voltage on the probe tip was swept from -95 V to +95 V in steps of 0.1 V. For every voltage step, the current collected by the probe tip was measured ten times, and the average was then computed by the probe software and stored. This measurement was repeated several times, for each radial position. Employing the IV-characteristic data, the probe software computed several plasma parameters using standard Laframboise theory for cylindrical probes \cite{Laframboise}. Parameters computed include the electron and ion density, $n_e$, $n_{+}$, the plasma potential, $V_P$, the floating potential $V_{fl}$, and the electron temperature $T_e$.

\section{Description of the numerical model}\label{sec:model}

A two-dimensional hydrodynamic model is used to solve the equations for the electron- and ion-fluid (in an
argon discharge), coupled to the dust fluid. In this study, we are concerned with the effect of changing discharge settings on the dust transport, and the effect of the modifications in the modified GEC cell on the plasma parameters. We therefore only consider situations where small numbers of dust particles are present in the discharge, so that the plasma parameters, and the forces acting on the dust, are not altered by the dust itself. However, the forces are still self-consistenly computed from the plasma parameters. We now proceed with a description of the solution for the plasma parameters.

\subsection{Solution scheme for the plasma.}
The continuity equation for the density $n_j$ for the electrons and ions ($j=e,ions$) is solved using a drift-diffusion approximation for the flux, ${\boldsymbol{\Gamma}}_j$:

\begin{equation}\label{eq:1}
\frac{\partial n_j}{\partial t} + \nabla \cdot {\boldsymbol{\Gamma}}_j = S_j, ~~~ {\boldsymbol{\Gamma}}_j = n_j{\mu}_j\boldsymbol{E} - D_j\nabla n_j,
\end{equation}

\noindent with ${\mu}_j$ the mobility and $D_j$ the diffusion coefficient. The sinks and
sources $S_j$ include electron-impact ionization and electron-impact excitation. The
electric field is found from the Poisson equation, 

\begin{equation}\label{eq:2}
{\nabla}^2 V = -\frac{e}{{\epsilon}_0}\left(n_{+} - n_e\right),~~~ \boldsymbol{E} = -\nabla V,
\end{equation}

\noindent with $n_{+}, n_e$ the ion and electron density, ${\epsilon}_0$ the permittivity of vacuum, and $e$ the electron charge. 

Since the argon ions are too massive to follow the instantaneous electric field, an effective electric
field is calculated to include the effect of ion inertia by solving $d{\boldsymbol{E}}_{eff}/dt = {\nu}_{m,+}(\boldsymbol{E}-{\boldsymbol{E}}_{eff})$, with ${\nu}_{m,+}$ the momentum transfer frequency for ion-neutral collisions. 

A similar set of equations is solved for the average electron energy density, $w = n_e {\epsilon}$, with ${\epsilon}$ the average electron energy.

\begin{equation}\label{eq:3}
\frac{\partial w}{\partial t} + \nabla \cdot {\boldsymbol{\Gamma}}_{w} = -e{\boldsymbol{\Gamma}}_{e}\cdot\boldsymbol{E}
+ S_{w}, ~~~ {\boldsymbol{\Gamma}}_{w} = \frac{5}{3}\left({\mu}_ew\boldsymbol{E} - D_e\nabla w\right).
\end{equation}

\noindent In the above, $-e{\boldsymbol{\Gamma}}_{e}\cdot\boldsymbol{E}$ is the Ohmic electron heating, which is the power input. The sinks, $S_w$, include electron
impact ionization and excitation. The ions are assumed to locally dissipate their energy, so that it is not necessary to solve a similar equation for the ions. Equations (\ref{eq:1}, \ref{eq:2}, \ref{eq:3}) are progressed in time on sub-RF timescales until the solution set $U(t)$ becomes periodic over a RF cycle to within a very small user-defined parameter; $(U(t) = U(t+{\tau}_{RF}))$. 

\subsection{Solution scheme for the dust fluid.}

\subsubsection{Dust particle charging}
A spherical dust particle with radius $a$ immersed in plasma absorbs electrons and ions (with mass $m_e$ and $m_{+}$ respectively) until in equilibrium the electron and ion currents balance. Due to the high electron mobility compared to ion mobility, the equilibrium dust charge becomes negative, $V(a) <0$. Using energy and angular momentum conservation, the ion and electron current can be calculated from OML theory \cite{OML} through

\begin{eqnarray}\label{eq:4}
I_{+} &=& 4\pi a^2 e n_{+}\sqrt{\frac{E_{+}}{2m_{+}}}\left[1-\frac{eV(a)}{E_{+}} + 0.1
{\left(\frac{eV(a)}{E_{+}}\right)}^2\left(\frac{{\lambda}_D}{l_{mfp}}\right)\right],\\
I_e &=& -4\pi a^2 en_e\sqrt{\frac{kT_e}{2\pi m_e}}\exp{\left(\frac{eV(a)}{kT_e}\right)}.
\end{eqnarray}

\noindent For typical dusty plasma experiments, charge-exchange collisions increase the ion current to the dust particles, making the equilibrium dust charge less negative \cite{OML-coll}. This is included in the above equations as the final term between square brackets in the ion current equation. In the equations, $T_e$ is the electron temperature, $k$ is Boltzmann's constant, ${\lambda_D}$ is the linearized Debye length, and
$l_{mfp}$ is the ion-neutral collision mean free path. $E_{+}= \frac{1}{2} m_{+} v^2 = \frac{4kT_{+}}{\pi} + \frac{1}{2}m_{+}u_{+}^2$ is the mean ion energy, with $T_{+} \approx T_{gas}$ the ion temperature \cite{barnes} and $u_{+}$ is the ion velocity found from the flux, equation (\ref{eq:1}). The dust charge number is found using a capacitor model: $eZ_D = 4\pi{\epsilon}_0aV(a)$. 

\subsubsection{Dust particle transport}
The forces acting on dust particles include \emph{gravity, the electrostatic force, ion drag, neutral drag}, and \emph{thermophoresis}.
The gravitional and electrostatic force are calculated from

\begin{equation}\label{eq:7}
{\boldsymbol{F}}_g = m_D\boldsymbol{g} = -\frac{4\pi}{3}a^3\rho g ~\hat{\boldsymbol{e}}_z,~~~{\boldsymbol{F}}_{E} =
eZ_D\overline{\boldsymbol{E}},
\end{equation}

\noindent where $\rho = 1510$ kg m${}^{-3}$ for melamine-formaldehyde particles, $g=9.81$ m s${}^{-2}$, and $\overline{\boldsymbol{E}}$ the
time-averaged electric field. The neutral drag and thermophoretic force can be calcuated using

\begin{equation}\label{eq:8}
\boldsymbol{F}_n = -\frac{4\pi}{3}a^2{\rho}_{gas}v_{gas}\boldsymbol{v}_D,~~~{\boldsymbol{F}}_{th} =
-\frac{32}{15}a^2\frac{{\kappa}_T\nabla T_{gas}}{v_{gas}},
\end{equation}

\noindent with $v_{gas}$ defined as the thermal velocity of the neutral gas, ${\rho}_{gas} = m_{Ar}n_{gas}$ the neutral gas mass
density for argon, $\nabla T_{gas}$ the temperature gradient of the background gas, and ${\kappa}_T = 0.0177$ W K${}^{-1}$ m${}^{-1}$ the thermal conductivity of argon at 300 K.

The ion drag force is calculated including the effect of ion scattering beyond the Debye length, the effect of charge-exchange collisions and the effect of ion drift through

\begin{equation}\label{eq:9}
\boldsymbol{F}_{id} = n_{+}m_{+}v\boldsymbol{u}_{+}\left({\sigma}_c(v) +
\pi{\rho}_C^2(v)\left[{\Lambda}(\tilde{v})+\mathcal{K}\left(\frac{{\lambda}_D}{l_{mfp}}\right)\right]\right),
\end{equation}

\noindent where $v$ is determined from the mean ion energy, ${\sigma}_c(v)$ is the OML collection cross section \cite{OML}, ${\rho}_C (v)$ the Coulomb radius, $\Lambda(\tilde{v}) $ 
the Coulomb logarithm for scattering beyond the Debye length \cite{Khrapak2002}, and $\tilde{v}$ is found from a fit for the total
energy in the Coulomb logarithm in order to take the effect of significant ion drift on the screening into account
\cite{SCEPTIC}. $\mathcal{K} (x) =
x\arctan(x) + \left(\sqrt{\frac{\pi}{2}}-1\right)\frac{x^2}{1+x^2} - \sqrt{\frac{\pi}{2}}\ln(1+x^2)$ is the collision
operator, used to calculate the effect of charge-exchange collisions on the ion drag \cite{IVLEV}. 

Assuming a force balance between the neutral drag and all other forces, a drift-diffusion type equation can be derived for the
dust flux,

\begin{equation}\label{eq:10}
{\boldsymbol{\Gamma}}_D = n_D
{\left(m_D{\nu}_{mD}\right)}^{-1}\left[\boldsymbol{F}_g + \boldsymbol{F}_{E} + \boldsymbol{F}_{th} + \boldsymbol{F}_{id}
\right].
\end{equation}

\noindent with ${\nu}_{mD}$ the dust-neutral momentum transfer frequency. This equation can be used to transport the dust in time. In this study, we are simply concerned with the differences in the net force, due to changes in the geometry of the modified GEC cell, as well as the effect of the discharge settings on the forces. 

\subsubsection{Geometry}

The cylindrical symmetry allows us to model half of the volume, approximated by a half-\emph{H}-shape, and the two spatial directions become the radial direction, with coordinate $r$ and the axial direction, with coordinate $z$. The biggest difference between the real and the model geometry is that in the model the top (grounded) electrode is solid, whereas in the experiment it is a hollow cylinder. The radial cutout plate, which provides radial confinement for the dust particles, is added in the model as an additional boundary condition for the potential on the inner part of the lower electrode, varying quadratically from the center of the electrode to the edge of the cutout. The numerical grid solution is 48 radial by 96 axial points. The vertical grid points are divided in three regions. Region I between the electrodes (2/3 of the points, giving a resolution of 3 cm/64 points = 0.5 mm/grid interval) and regions II and III next to the electrodes (both with 1/6 of the grid points).

Once the plasma parameters are converged, the solutions on the grid are used to compute the dust charge, using the OML charging currents from equations \ref{eq:4} and 5. Once the dust charge is determined, the electrostatic force and the ion drag force are computed. The neutral gas temperature is solved with an energy conservation scheme, which in this study only involves the heating by local energy deposition of ions and the thermal conduction by the gas. Once the temperature gradient is resolved, the thermophoretic force can be calculated. The sum of these forces thus gives the net force that would act on dust particles suspended in the plasma.

\section{Results for global discharge parameters}\label{sec:globalresults}

The global characteristics can be measured by the absorbed power and the natural DC bias on the powered electrode. If our model is to correctly represent the actual experiment, the DC bias and the total power absorbed in the discharge should be similar to the experiment for given discharge settings. We therefore first compare these two easily determined global parameters.

\subsection{Natural DC bias.}

In asymmetric discharges, such as those found in the GEC cell, the number of electrons collected by the powered electrode per RF cycle during the positive phase of the cycle is unequal to the amount of ions collected during the negative phase. Therefore, the surface of the powered electrode acquires a negative charge, corresponding to a negative potential, called the (natural) DC bias. As a first approximation \cite{Song}, it only depends on the relative surface of the \emph{plasma facing grounded area} ($A_g$) and the
\emph{plasma facing powered area} ($A_d$), and the amplitude of the driving potential, or the root-mean-square value, $V_{rms} = V_0/\sqrt{2}$,

\begin{equation}\label{eq:14}
\frac{V_{DC}}{V_{rms}} = \sqrt{2}\sin\left[\left(\frac{\pi}{2}\right)\frac{A_d-A_g}{A_d+A_g}\right]. 
\end{equation}

\begin{figure}[htbp]
\begin{center}
\includegraphics[width=0.55\textwidth]{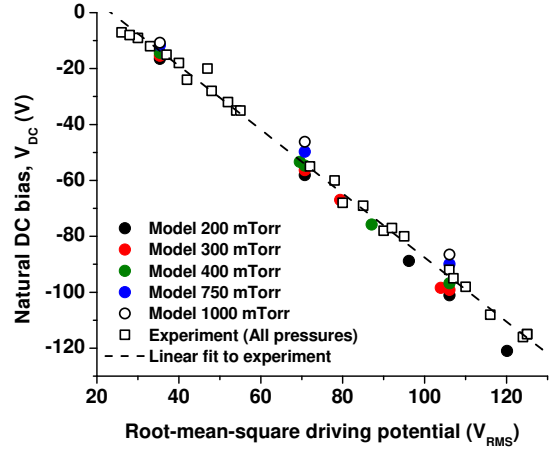}
\caption{The experimentally determined DC bias (open squares) together with a linear fit to the data (dashed line).
The circles are results from the model, showing good agreement with the experiment as well as a dependence on the
pressure in agreement with the literature.}\label{fig:2}
\end{center}
\end{figure}

\noindent Figure \ref{fig:2} shows the experimentally determined DC bias versus the root-mean-square
driving potential, as well as modeling results for various pressures. The dashed line is a linear fit to the experimental
data. The results from the model, calculated by adjusting the DC potential until the electron and ion current to the powered electrode balance, fall on the line, with the DC bias becoming more negative for lower pressures, which is in agreement with the literature \cite{GEC3, DCmodel}, and discussed below in section \ref{sec:summary}.

\subsection{Absorbed input power.}

\begin{figure}[htbp]
\begin{center}
\includegraphics[width=0.55\textwidth]{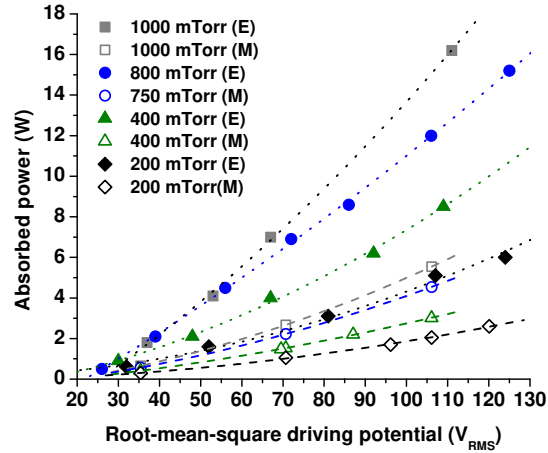}
\caption{The dissipated power, $P_{dis}$, measured experimentally (closed symbols, 'E') and the total absorbed power, $P_{abs}$, computed by the model (open symbols, 'M') and second order polynomial fits versus driving potential, for different neutral pressures. The fits are very reliable ($R^2>0.99$), which shows that $P_{abs},P_{dis} \propto C(V-V_0)^2$.}\label{fig:3}
\end{center}
\end{figure}

The \emph{electrode power} is experimentally determined by measuring the (root-mean-square) potential applied to the powered electrode,
$V_{rms}$ and the current reaching the electrode, $I$. Assuming that for the given powers and pressures examined the plasma acts much like a simple impedance, the total power dissipation in the plasma is then $P_{dis} = V_{rms} I$ \cite{BernardThesis}. In the model, the total power \emph{absorbed by the charged species} is calculated from the volume integral of the Ohmic heating term, $P_{abs} = \int\int\int\left({\boldsymbol{J}}_e\cdot\boldsymbol{E} + {\boldsymbol{J}}_{+}\cdot\boldsymbol{E}\right) dV$. The resulting power versus driving potential plot is shown in figure \ref{fig:3}. The dotted and dashed lines are second order polynomial fits to the data, with coefficients of determination, $R^2$, greater than 0.99, which implies the fits are highly reliable. We observe that the modeled absorbed power $P_{abs}$ and the measured dissipated power $P_{dis}$ differ by a factor of 2-4, which might indicate a difference in the model and actual plasma conductivity, as discussed in section \ref{sec:summary}.

\section{Results for local parameters}\label{sec:localresults}

Dust clouds will be levitated in the modified GEC cell under the influence of different forces. We consider the electrostatic force, which depends on the dust charge and the electric field, and the ion drag force, which depends on the drift velovity of the ions, but mainly on the ion density in the discharge. The equilibrium position of dust particles is determined by the position where the net force acting on the dust particles vanishes. 

In the vertical direction, this balance is mainly determined by the electrostatic force and gravity (depending on the size of the particles, the ion drag might play an important role in the force balance). Since the vertical electric field required to balance gravity, $E=m_Dg/q_D$, is rather large for micrometer sized melamine-formaldehyde particles, the particles will in general levitate near, or in the sheath above the powered electrode.

In the radial direction, the electric field due to the plasma potential is generally relatively weak. In order to create a potential well in which the particles will be trapped, an additional electric field is created by the cutout in the lower electrode. This additional electric field causes an additional radial force pulling the particles inward. Usually a balance between this electric field and the inter-particle interaction is assumed to determine the radial equilibrium of the particles. However, it has been shown in micro-gravity experiments that the ion drag force is responsible for the formation of a dust-free void in the plasma bulk, and has thus been shown to have a rather large magnitude. It is therefore not unlikely that the ion drag force plays an important role in the radial force balance.

The probe measurements were taken at the midplane between the two electrodes, which is above the equilibrium levitation height for most particle sizes, however, the profile behavior for varying discharge parameters will still give a good idea of the effect of these parameters on the forces acting on dust particles suspended in the discharge. These measurments can not only be used to compare the profiles obtained from our model with the experiment, but also to show differences between the obtained results and similar results reported for the original GEC reference cell. Since we were only able to perform Langmuir probe measurements at different radial positions, we can not make any statements on the change of the forces with height. Therefore, we mainly focus on the radial forces acting on dust particles that would be levitated in the discharge.

In this section the different radial plasma profiles obtained with the Langmuir probe at different applied powers, and at pressures of 200 mTorr and 400 mTorr, are compared to similar profiles obtained with the model at the same driving potential, determined from figure \ref{fig:3}. This means, for instance, that the experimental results at 400 mTorr and 6 Watts, which are run at roughly $V_{rms}$ = 90 V are compared to a model run at the same pressure and driving potential, which is roughly 2-2.5 Watts of Ohmic heating. In the figures, the labels in the legends refer to the pressure and the power, so that $200-6$ refers to results from a discharge at 200 mTorr and 6 Watts of power. 

\subsection{Dust charge.}

\begin{figure}[htbp]
\includegraphics[width=0.5\textwidth]{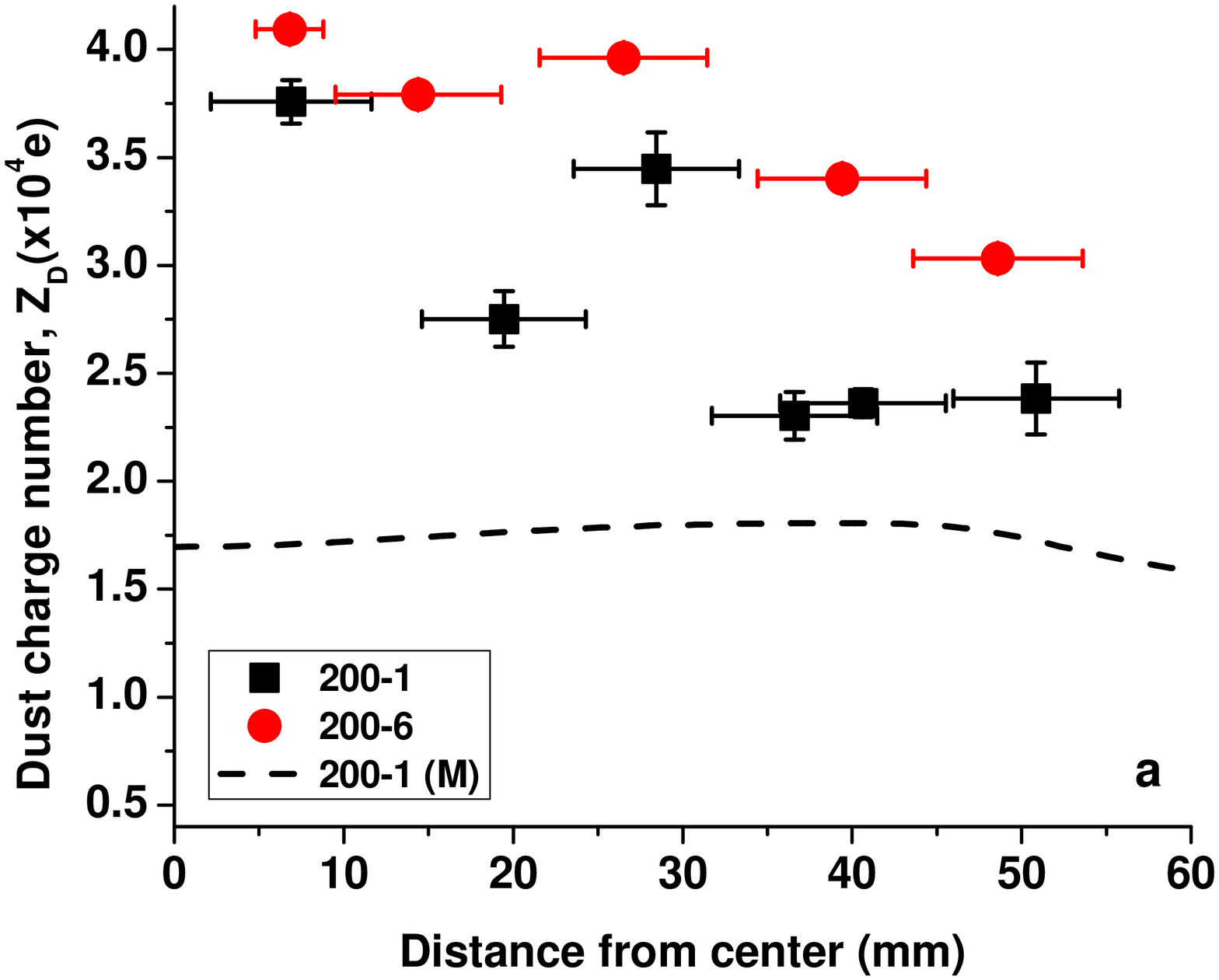}
\includegraphics[width=0.5\textwidth]{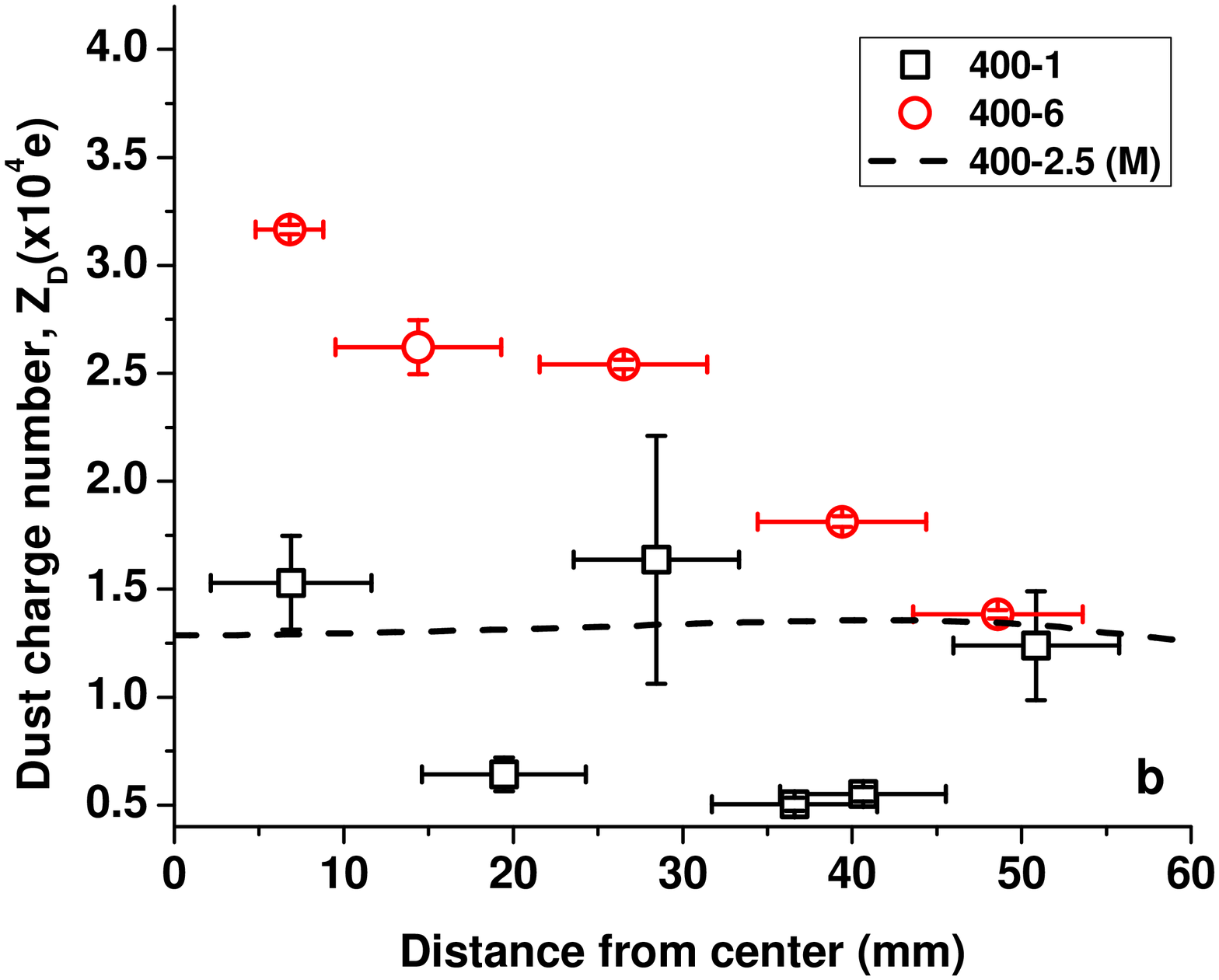}
\caption{The surface charge for particles with a radius of 4.45 $\mu$m. The results for 200 mTorr are shown on the left (a), the results for 400 mTorr on the right (b). The dashed lines represent model results for 200 mTorr at 1 Watt, and 400 mTorr at 2.5 Watts.}\label{fig:4}
\end{figure}

The equilibrium charge on a spherical particle depends on the size of the particle and the electron temperature. The Langmuir probe determines the local electron temperature, as well as the floating potential. The electron temperature determines the dust charge via $Z_D = -\frac{1}{2}T_e(eV)\ln{\left(\frac{m_{Ar}}{4\pi m_e}\right)}$ \cite{Hutchinsonbook}, while the dust charge number can be determined from the measured floating potential through $Z_D = V_{fl}(4\pi{\epsilon}_0a/e)$. Because of the larger error in the electron temperature obtained from the Langmuir probe, we use the second approach. Figure \ref{fig:4} shows the dust charge number determined from the measured floating potential for particles with a radius of 4.45 $\mu$m, which are often used in experiments in the modified GEC cell, at different radial positions above the lower electrode, as well as results for corresponding cases obtained with the model. 

The dust charge profiles differ significantly for the low power (1 Watt) and high power (6 Watts) case. For low power the dust charge shows a strong dip around 20 mm for both pressures, followed by a peak at 30 mm from the center, whereas at high power, the dip is insignificant and the dust charge decreases smoothly outwards. It is also clear that the dust charge decreases for higher pressure, consistent with our model results, and increases slightly with increasing power. A similar reduction of the dust charge with increasing pressure was also reported in \cite{ReducedCharge}. The difference between the model and the Langmuir probe data is roughly a factor of 2, which we expect has to do with the ion-neutral collisions. (See the next section and a discussion in section \ref{sec:summary}.) 

\subsection{Plasma densities.}

\begin{figure}[htbp]
\begin{center}
\includegraphics[width=0.5\textwidth]{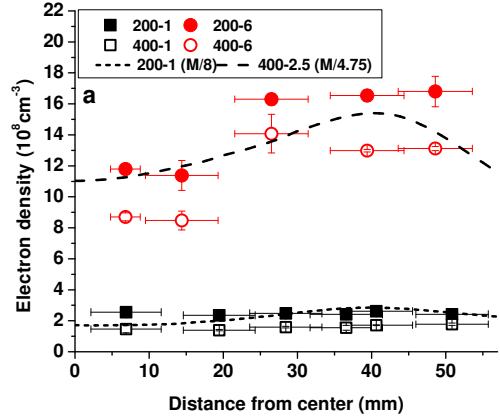}
\caption{The electron densities measured in the discharge (symbols) and two corresponding model solutions (dashed/dotted lines). Note that the model values have been divided by arbitrary factors, showing that the absolute value of the electron density predicted by the model is much higher.}\label{fig:5}
\end{center}
\end{figure}

\begin{figure}[htbp]
\includegraphics[width=0.5\textwidth]{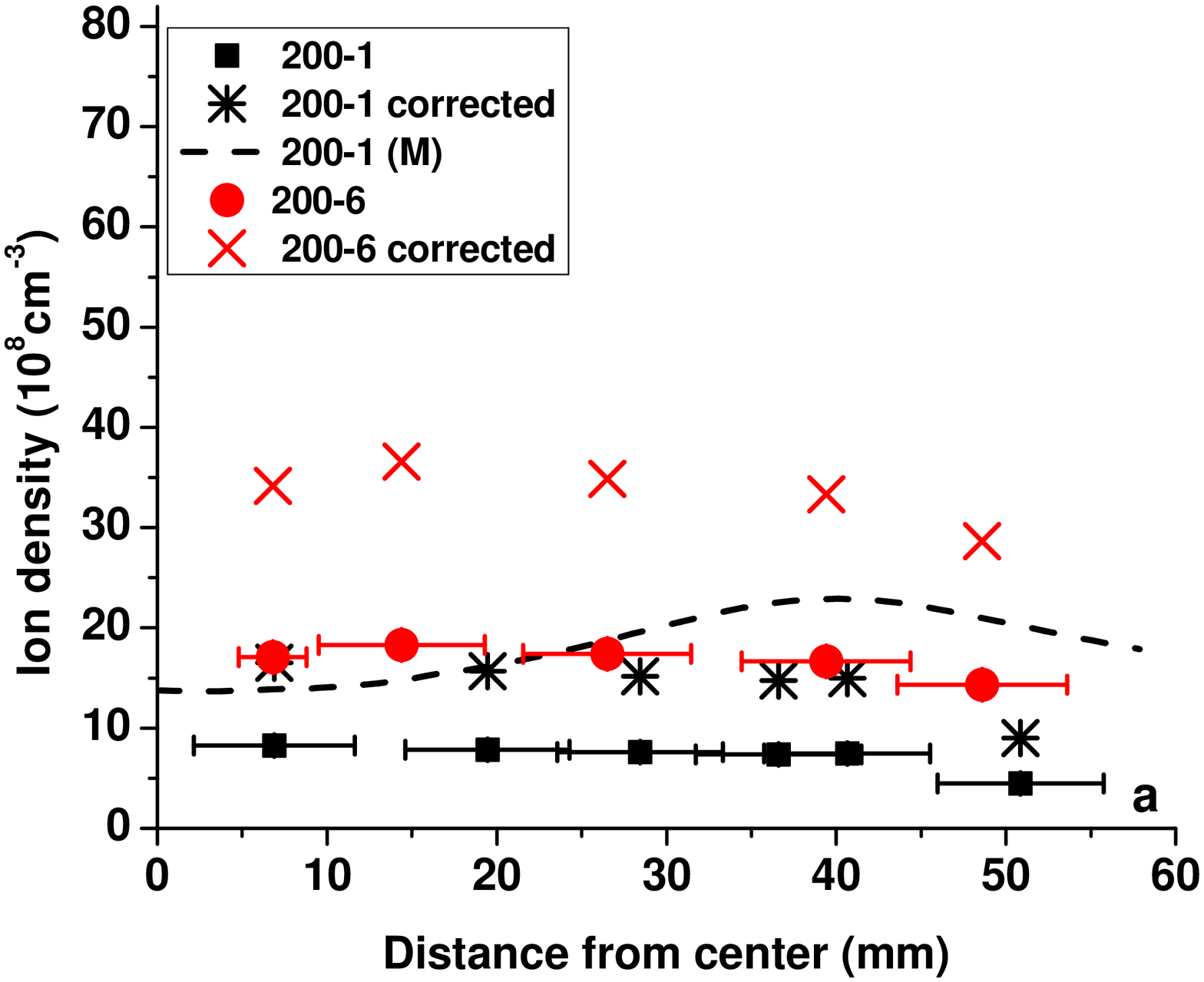}
\includegraphics[width=0.5\textwidth]{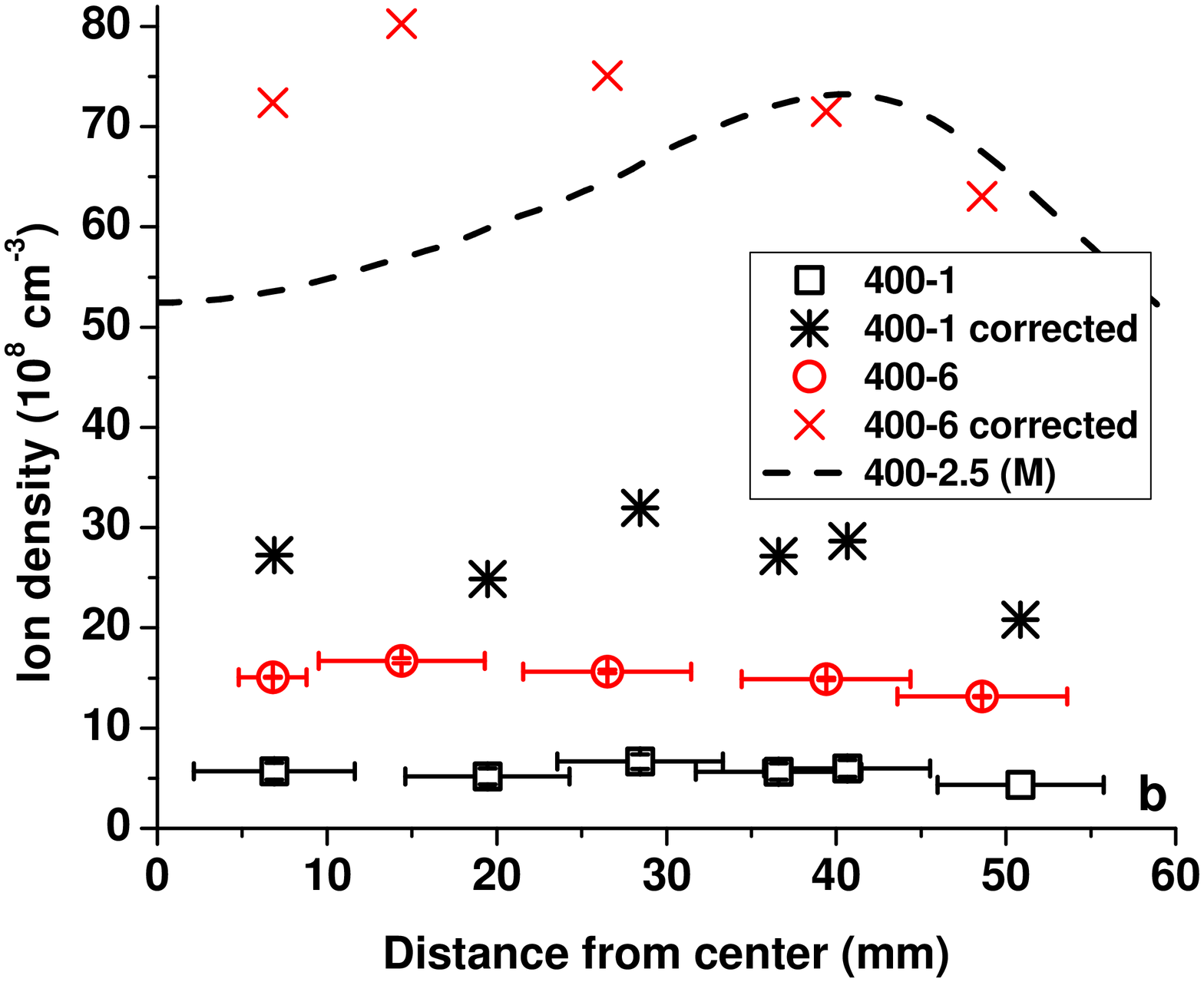}
\caption{The Ar${}^{+}$ densities at 200 mTorr on the left (a) and at 400 mTorr on the right (b) measured in the discharge (symbols) and the model solutions indicated by the dashed lines. The corrected ion densities are explained in section \ref{sec:summary} and have to do with ion-neutral collisions.}\label{fig:5b}
\end{figure}

The electron and ion densities in the experiment are shown in figure \ref{fig:5} and figure \ref{fig:5b}, respectively, together with model results. The general trend for the plasma densities is to increase with the absorbed power, however, the response of the ions and electrons to a change in neutral gas pressure is different.

We observe that the electron density increases roughly 6-fold when the dissipated power increases 6-fold, which might indicate that $n_e \propto P_{abs}$, even though there are only two values of the power considered here. The electron density decreases with neutral pressure, which is a result quite different from other results reported in the literature. The shape of the radial electron density profiles corresponds very well with the profiles obtained in the model, however, the absolute value of the electron density is much higher in the model. We also see that the difference between the model and the experiment changes with neutral pressure. This decrepancy might have to do with the decrepancy we observed between the absorbed and dissipated power. This will be further discussed in section \ref{sec:summary}.

The ion density increases with power, but by a factor 2-3 rather than a factor of 6. For increasing pressure, we see that the ion density decreases, but only by a small fraction. The ion density profile peaks in the center and falls off to the edge of the electrode. The model profiles are similar to the electron density profiles, having a peak off-axis. Furthermore, the model ion density is much higher than the experimentally obtained ion density.

A caveat is in place here, however. It has been shown in the past that the ion density is underestimated by ordinary Langmuir probe data in the original GEC cell at pressures above 50 mTorr \cite{GEC4}. It was shown that for 50 mTorr $<P_{gas}<$ 250 mTorr, the ion density is roughly a factor of 2 higher than the value obtained from a Langmuir probe, whereas for $P_{gas}>$ 250 mTorr, it becomes a factor $R_{probe}/l_{mfp}$ higher, with $R_{probe}$ the radius of the probe-tip and $l_{mfp}$ the ion-neutral momentum transfer mean-free path. At 400 mTorr, this results in a factor of 5. We have added these \emph{corrected} ion density profiles in the graphs. They are reasonably in agreement with the model results. 

Finally, for all powers and pressures, the ion density is higher than the electron density, so that the modified GEC cell seems to be strongly electropositive, which might indicate that a clear quasi-neutral bulk does not exist for these discharge settings. This is also found with the model, but the difference between the electron and ion density is much smaller.

\subsection{Plasma potential.}

\begin{figure}[htbp]
\includegraphics[width=0.5\textwidth]{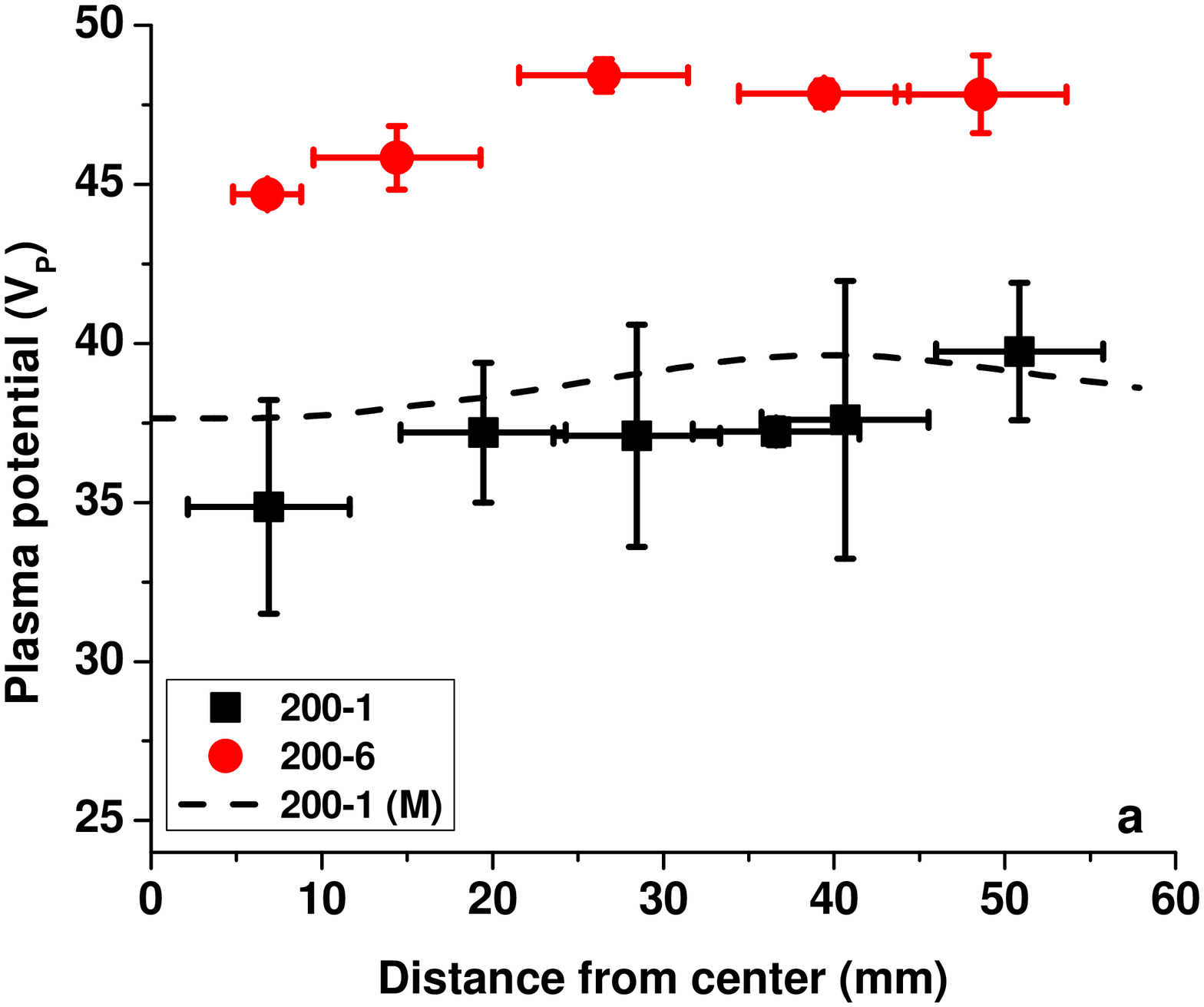}
\includegraphics[width=0.5\textwidth]{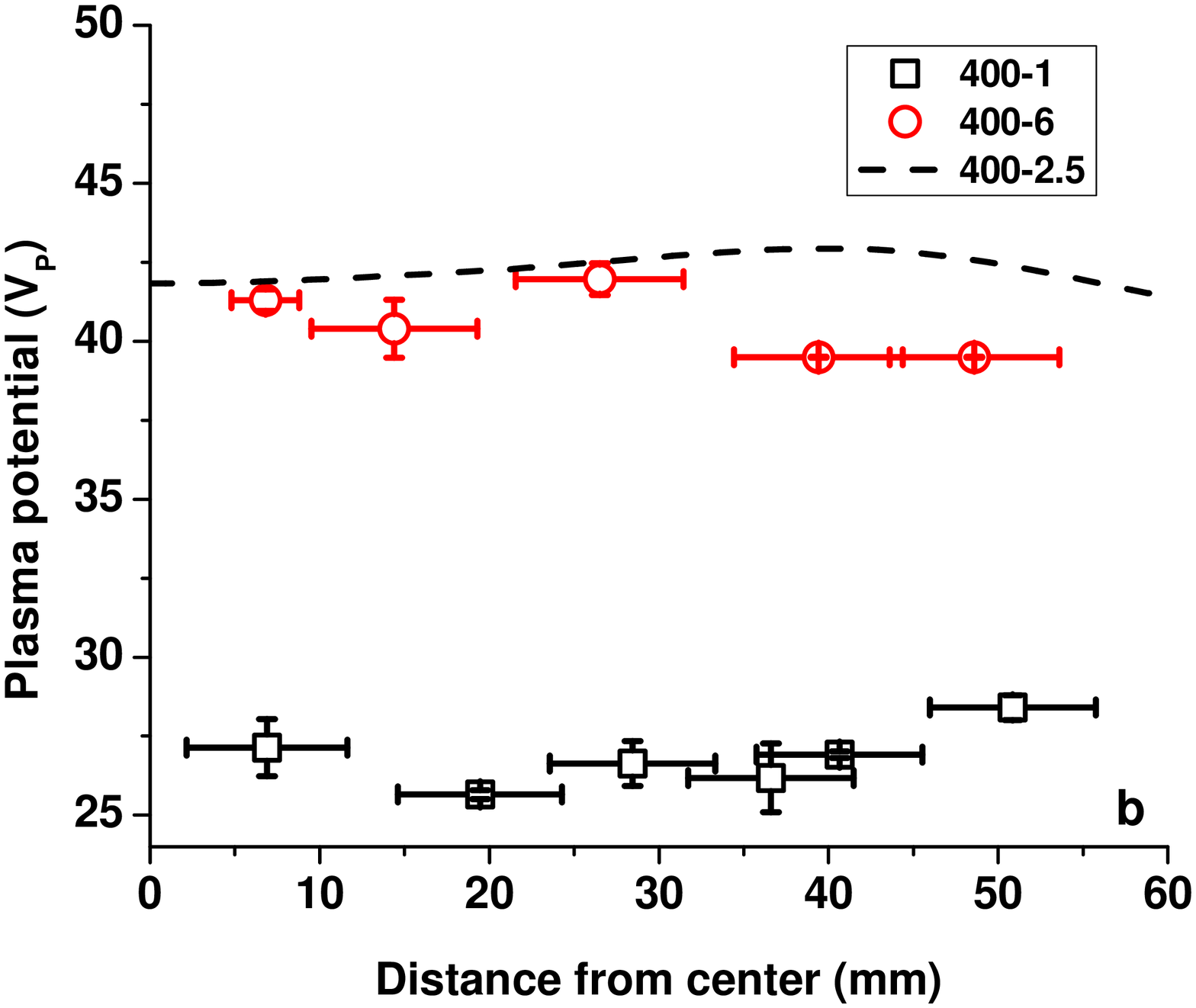}
\caption{The plasma potential at 200 mTorr (left, labeled a) and 400 mTorr (right, labeled b).}\label{fig:6}
\end{figure}

Figure \ref{fig:6} shows the measured plasma potential profiles as well as profiles found from the model, for 200 mTorr and for 400 mTorr. We see that the plasma potential shows the same behavior as the plasma densities, i.e. an increase with power, but a decrease with pressure. The model results in this case are close to the measured profiles, even though the model always results in an off-axis maximum in the plasma potential, related to the off-axis maximum in the plasma densities obtained in the model.

Although the profiles are relatively flat, we still see a distinct difference between the low pressure (200 mTorr) and high pressure (400 mTorr) case, namely that the former has a positive radial derivative everywhere, while the latter has a negative radial derivative in the inner region of the discharge. This means there is a small electric field pointing \emph{inward} for lower pressure, and hence a small electric force pointing \emph{outward} for negatively charged particles, whereas the opposite is true at higher pressures. Therefore, the electrostatic force seems to add to the radial confinement at higher pressures, but acts against the confinement at lower pressures, at least in the plane between the electrodes.

\section{Forces acting on the dust}\label{sec:forces}

Even though we do not have any data that would allow us to show any axial (vertical) force dependence on the discharge parameters, we can use the observed dust charge to see how large the electric force would have to be to balance gravity. For MF particles with $r_D = 4.5 \cdot{10}^{-6}$ m and $\rho$=1510 kg m${}^{-3}$, we find that the force of gravity equals $F_g = m_Dg =$ 5.65 $\cdot {10}^{-12}$ N. Using 15.000e as the minimum dust charge from figure \ref{fig:4}, we find that the upper bound for the vertical electric field required to balance gravity is $E_{z,max} = m_Dg/Q_D\approx$ 2400 V/m. This is roughly 10 times the value of tyical ambipolar electric fields, which shows that the particles will be suspended well below the midplane of the electrodes, towards the lower sheath.

In the radial direction, we are interested in the electrostatic force and the ion drag force. The Debye length, ${\lambda}_D\approx{\lambda}_{+} = 69\sqrt{T_{+}/n_{+}}$ can be determined from the obtained ion density. The Coulomb logarithm, $\Lambda$, can then be determined with the knowledge of the dust charge. The radial electric field can be determined from the gradient in the plasma potential, which, together with the ion density gradient, results in the ion drift velocity from equation \ref{eq:1}. Knowing all these parameters, the ion drag can be determined. The electrostatic force is determined from the dust charge and the determined electric field. 

\begin{figure}[htbp]
\includegraphics[width=0.5\textwidth]{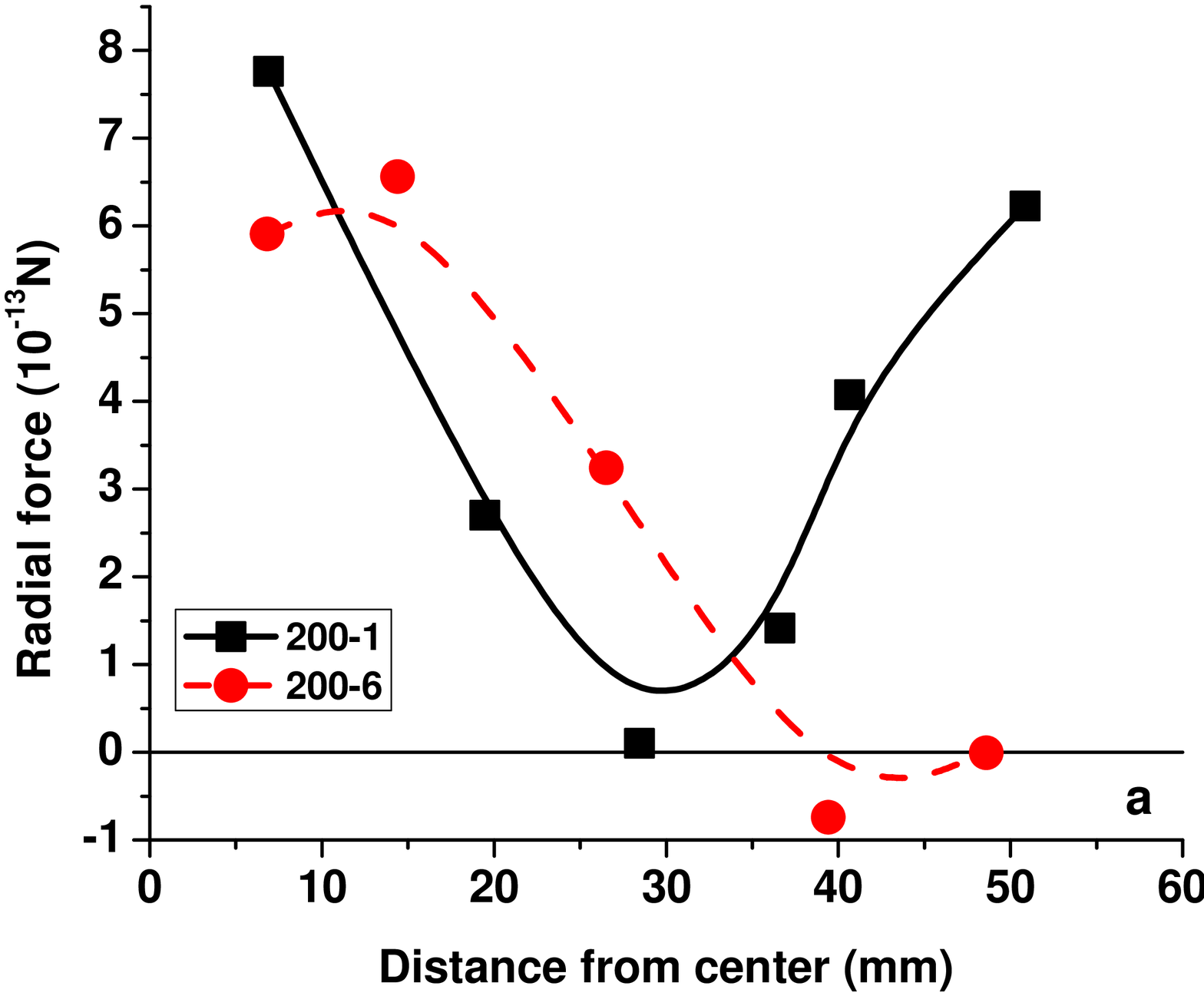}
\includegraphics[width=0.5\textwidth]{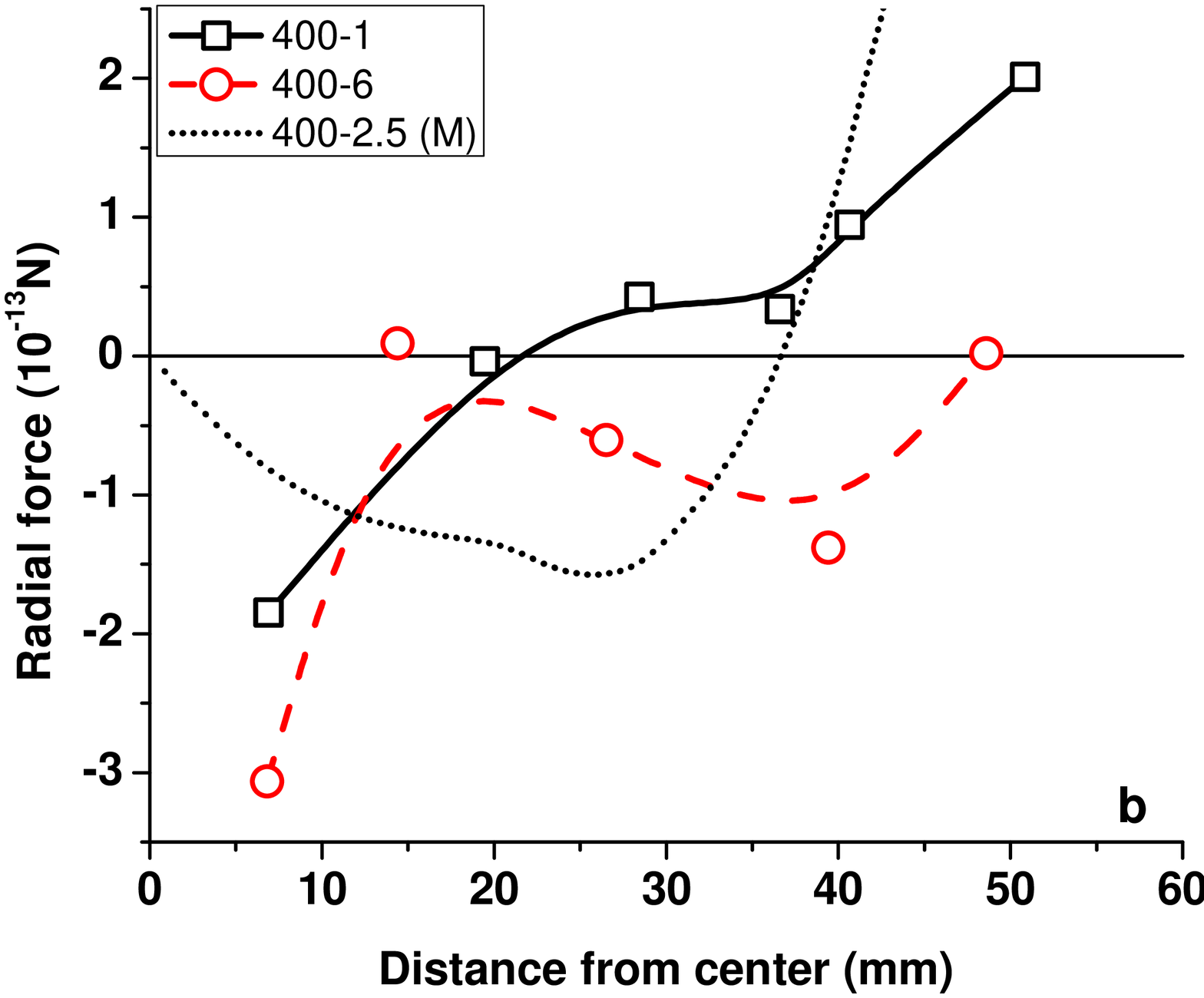}
\caption{The net radial force at the midplane calculated from the Langmuir probe data for 200 mTorr on the left (a), and 400 mTorr on the right (b).}\label{fig:7}
\end{figure}

The resulting net force obtained from the Langmuir data in this way is shown in figure \ref{fig:7} for 200 mTorr (a) and 400 mTorr (b). For the ion density we used the densities obtained by the Langmuir probe corrected for ion-neutral collisions. We see that for high pressure the force is directed inward near the center, whereas for lower pressure the force is directed outwards almost everywhere, except near the outer edge of the electrodes.

The radial force profile obtained from the model is similar for both pressures, since no electric field reversal is apparent in the model results. The model force profile at 400 mTorr and 2.5 Watts is shown on the right in figure \ref{fig:7}. Both the magnitude as well as the shape are in good agreement with the experimental results. This indicates that the corrected ion density, obtained with the Langmuir probe, as well as the measured plasma potential profile (and its derivative) are reliable, since the model equations for the forces are generally believed to be accurate.

\section{Discussion and outlook}\label{sec:summary}

In this section we discuss our results, point out the differences between the current results and the original GEC cell results reported in the literature, and discuss the possibilities for future work. We start our discussion with the results obtained for the global parameters of the modified GEC cell used for dusty plasma experiments.

\subsection{Natural DC bias.}

A first important conclusion for the model is that the difference between the model and the actual experimental geometry (i.e. the
hollow grounded electrode vs. a solid electrode), thus the difference in area
interacting with the plasma, is negligible, meaning that the model geometry is an excellent approximation to the real geometry. When we measure the slope of the fit in figure \ref{fig:2} and use the result in equation \ref{eq:14} we obtain a ratio of $A_g/A_d= 4$. The surface of the
powered area is simply $A_d = \pi r_{RF}^2 \approx92 $ cm${}^{-2}$. The total available grounded area in the experiment, $A_g \approx 3500$ cm${}^{-2} \approx 40 A_d$. Apparently, the grounded surface area interacting with the plasma (\emph{plasma facing}) is much smaller than the total available grounded surface.

Since the floating potential is directly related to $T_e$, the dependence of $T_e$ on the pressure should explain the behavior of the DC bias with pressure. The global balance that determines the electron temperature is the balance between the creation of particles (through electron impact ionization, with a ionization constant $K_{iz}(T_e)$ (m${}^{3}$s${}^{-1}$)) and the losses of particles (through the flux of particles to the walls at the Bohm velocity, $u_B=\sqrt{kT_e/m_{+}}$). For a cylindrical discharge with radius $R$, height $H$, gas density $n_{gas}$, and electron density $n_e$ this balance can be written as \cite{Liebermanbook}

\begin{equation}
K_{iz}n_{gas}n_e\pi R^2H=\left(2\pi R^2 a + 2\pi RHb\right)n_eu_B,
\end{equation}

\noindent with $a=0.8/\sqrt{4+R/l_{mfp}}$ and $b=0.86/\sqrt{3+h/2l_{mfp}}$. Here $l_{mfp}=1/n_{gas}{\sigma}_{in}$ the mean-free path for ion-neutral collisions (including both elastic scattering and charge-exchange collisions, ${\sigma}_{in} \approx {10}^{-18}$ m${}^{2}$). This form of the particle balance is valid for $P_{gas} \approx 100$ mTorr. Reordering the terms, we find

\begin{equation}\label{eq:balance}
\left(n_{gas} d_{eff}\right)^{-1} = \frac{K_{iz}(T_e)}{u_B(T_e)},
\end{equation}

\noindent where $d_{eff}= 0.5 RH/(aR+bH)$. We obtained a good approximation for $n_{gas}d_{eff}$ in terms of the pressure, $P_{gas}=n_{gas}k_BT_{gas}$, with $R=$ 0.1 m and $H=$ 0.2 m for our modified GEC cell:

\begin{equation}\label{eq:ngasdeff}
n_{gas}d_{eff}(n_{gas}) \approx \frac{\sqrt{k_BT_{gas}}}{5}\left(\frac{P_{gas}}{k_BT_{gas}}\right)^{3/2},
\end{equation}
 
\noindent whereas a numerical approximation of the ionization coefficient is given by \cite{PasschierThesis}:

\begin{equation}
K_{iz} = 4.8
\cdot{10}^{-17} \left(\frac{3}{2}T_e (eV) - 5.3\right)\exp{\left(\frac{-4.9}{\sqrt{\frac{3}{2}T_e (eV)-5.3}}\right)}.
\end{equation}

\noindent Using a Taylor expansion around $T_e\approx$ 5 eV, which is the electron temperature we expect in the modified GEC cell, we can find an approximation to the ionization coefficient up to second order, which is correct within 4\% for 4 eV $<T_e<$ 7 eV:

\begin{equation}
K_{iz}(T_e)\approx 3.9\cdot{10}^{-18}+7\cdot{10}^{-18}\left(T_e-5\right)+3.2\cdot{10}^{-18}\left(T_e-5\right)^2.
\end{equation}

\noindent With $u_B(T_e) \approx 1550 \sqrt{T_e (eV)}$ a good approximation for $K_{iz}(T_e)/u_B(T_e)$ valid for 4 eV $<T_e<$ 7 eV is given by

\begin{equation}\label{eq:approxfrac}
\frac{K_{iz}(T_e)}{u_B(T_e)} \approx 2.1\cdot{10}^{-21}\frac{\left(T_e-3.9\right)^2}{\sqrt{T_e}}.
\end{equation}

Using equation (\ref{eq:ngasdeff}) and(\ref{eq:approxfrac}) in equation (\ref{eq:balance}) and taking the logarithm on both sides we find:

\begin{equation}
\log\left(\frac{P_{gas}}{{10}^{-2/3}}\right) = \log \left(T_e^{-2/3}\right)~\rightarrow T_e  = \frac{0.1}{P_{gas}^{3/2}}.
\end{equation}

\noindent It follows that the electron temperature decreases with increasing pressure, but the change in electron temperature for 200 mTorr $<P_{gas}<$ 400 mTorr is small. This means that the floating potential at a given driving potential becomes less negative for increasing pressure, but that the effect is small, which is also the result obtained with our model.

The above behavior for the DC bias is very similar to the behavior in the original GEC reference cell, and was found both in modeling \cite{GEC3}, as well as in many experimental studies, for instance in \cite{Foest}. We therefore conclude that the modifications in the modified GEC cell studied here have no significant effect on the global behavior of the floating DC bias.

\subsection{Dissipated and absorbed power.}

\begin{figure}[htbp]
\includegraphics[width=0.5\textwidth]{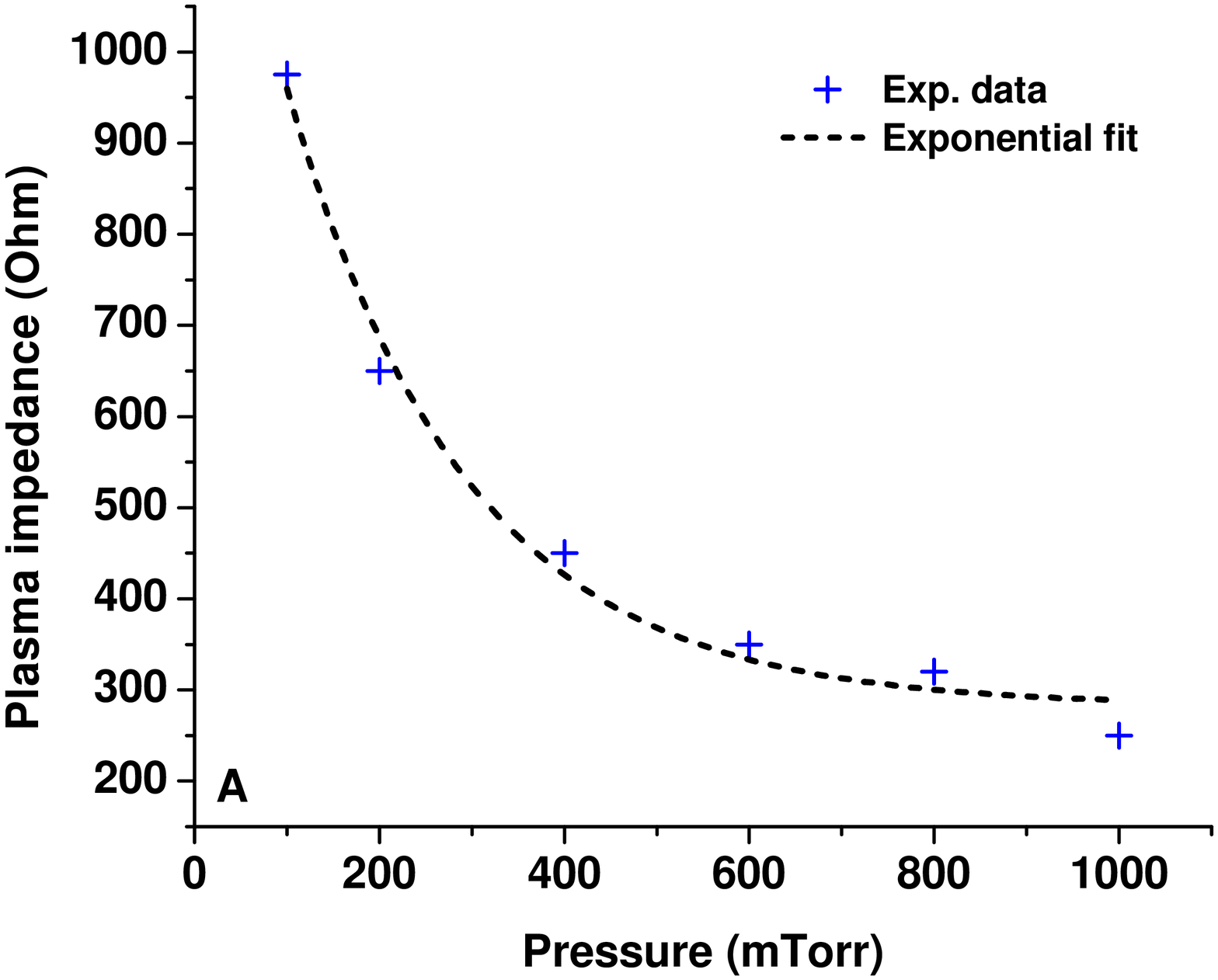}
\includegraphics[width=0.5\textwidth]{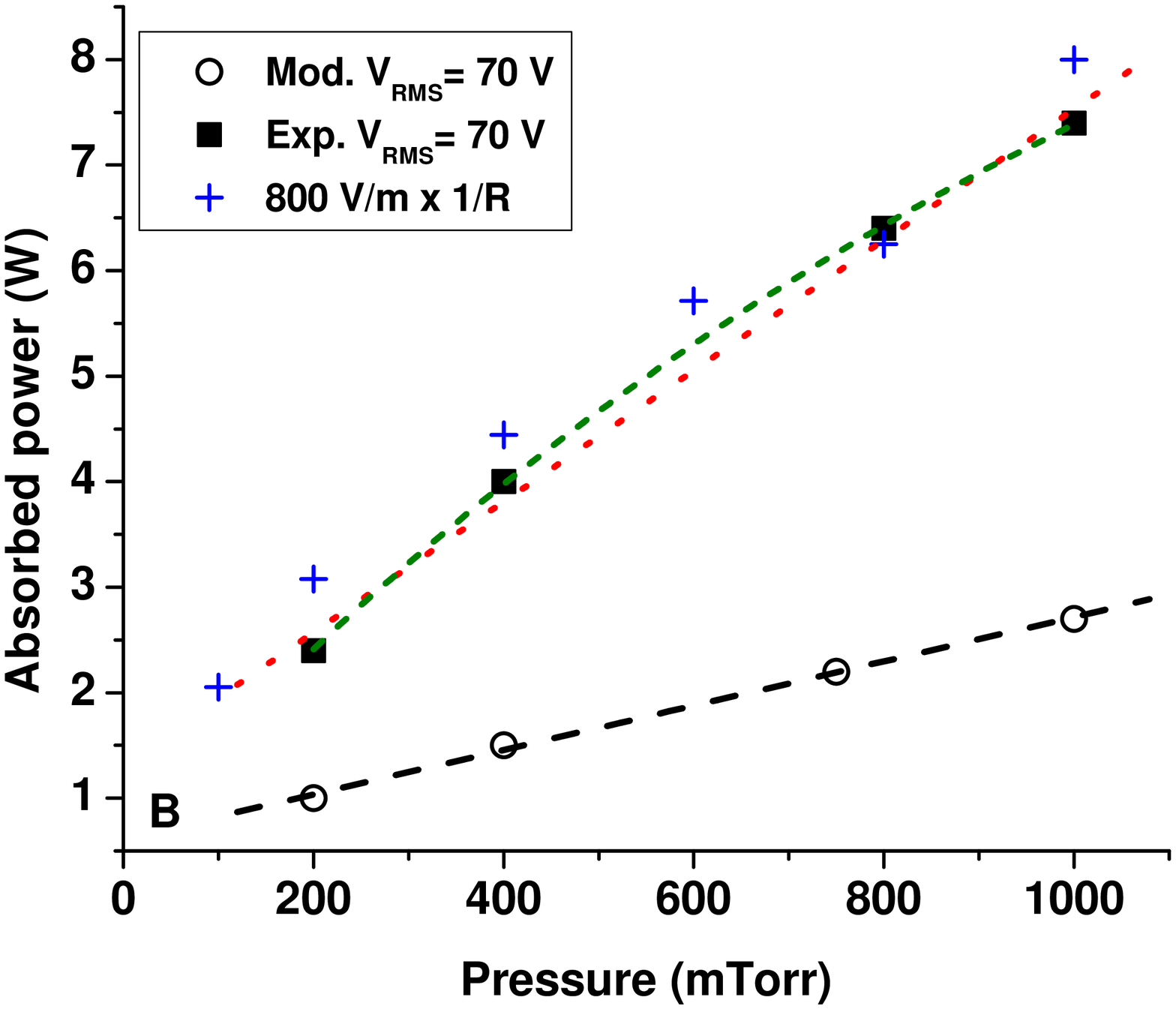}
\caption{A) A plot of the impedance determined in the modified GEC cell. B) The dissipated power at $V_{RMS} = 70$ V for the experiment (black squares), together with a reasonable linear fit (red, dotted line) and a much better exponential fit (green, dashed line). The absorbed power in the model is represented by the circles, an excellent linear fit is represented by the black, dashed line.}\label{fig:8}
\end{figure}

By measuring the plasma current and the applied potential, the plasma impedance was determined at different pressures. The results are shown in figure \ref{fig:8} on the left. The data is perfectly fitted by an exponentially decreasing impedance. Since the conductivity, $\sigma$, of the plasma is the reciproke of the impedance, the root-mean square absorbed power, $P_{dis}=\sigma E_{RMS}^2$ can be determined, if we assume a certain electric field. Figure \ref{fig:8} on the right shows the dissipated power assuming an electric field, $E_{RMS}=800$ V m${}^{-1}$ (represented by the blue crosses). 

The power dissipated in the experiment at $V_{RMS} = 70$ V, determined form figure \ref{fig:2} is shown by the black squares. We see that the assumption of a constant root-mean-square electric field of 800 V m${}^{-1}$ at this driving potential is acceptable. Included in the figure is a linear fit to the experimental data, represented by the red dotted line, which is reasonable, and an exponential fit, represented by the green dashed line, which is very good, with $R^2>0.99$.

Finally, the absorbed power calculated by the model at 70 V is represented by the open circles. The absorbed power can be perfectly fitted by a linear dependence on the pressure. In our model, we use an approximation for the pressure dependence of the electron-neutral momentum-transfer frequency, and hence the electron mobility \cite{PasschierThesis}:

\begin{equation}
{\mu}_{e,me} = 0.3\left(\frac{1000}{P_{gas}(mTorr)}\times\frac{T_{gas}}{273}\right).
\end{equation}

\noindent Since $\sigma = en_e{\mu}_e$, and ${\mu}_{e,m} \propto P_{gas}^{-1}$ in our model, we must have $n_e \propto P_{gas}^2$, which is consistent with the faster than linear increase in electron density found with our model.

Clearly, there is a difference between the actual electron-neutral momentum transfer frequency (and hence mobility, resistivity and conductivity) and the approximation used in the model. This can explain the difference in the electron densities obtained with the model and the ones measured in the modified GEC cell. Obviously, the determination of a better approximation for the electron mobility in the model will be part of future work.

The fact that the dissipated and absorbed power are best fit by a second order polynomial of the driving potential, rather than just a purely quadratic dependence, is due to the non-vanishing electric field in the electropositive discharge present in the modified GEC cell. We can write the electric field as a time-averaged part and an oscillating part ($E = E_0 + \tilde{E}$), so that we find for the time-averaged absorbed power

\begin{equation}\label{eq:splitE}
\overline{P} \sim \sigma \overline{E^2} dV \propto E_0^2 + 2E_0\tilde{E} + \overline{\tilde{E}^2}.
\end{equation}

\noindent In a symmetric discharge, the average electric field, $E_0$, is very small. Since $\tilde{E} \propto V(t) \propto V_{RF} \sin({\omega}_{RF}t)$, we would simply have $\overline{E \cdot E} \sim 0.5 |\tilde{E}|^2 \propto 0.5 V_{RF}^2$. However, in the modified GEC cell, the time averaged space charge does not vanish, as seen in the results for the plasma densities in figure \ref{fig:5} and figure \ref{fig:5b}, and the time averaged electric field is considerable. 

The behavior obtained for the absorbed power is similar to the observed pressure dependence of the power in the original GEC cell \cite{Sobolewski}. The modifications in our modified GEC cell therefore do not significantly change the global behavior of the absorbed power strongly. We now proceed by discussing the local parameters.

\subsection{Local parameters.}

The observed dust charge number lies in the range $5\cdot10^3<Z_D<4.2\cdot10^4$, increases with power and decreases with pressure. The last observation can in part be attributed to the effect of the pressure on the electron temperature, similar to the effect this has on the DC bias. A measurement of the dust charge, using single dust particle-dust particle interactions in the originial GEC cell at low pressures \cite{Konopka} indicated that the dust charge number for similar sized particles was in the ${10}^4-{10}^5$ range for screened charged particles, whereas it was always lower than $4\cdot{10}^3$ for unscreened particles. The first assumption seems to be consistent with our measurements. 

In \cite{Melzer} the dust charge was shown to decrease with pressure at pressures similar to our study. This was determined from the corresponding decrease in coupling parameter, $\Gamma$, for increasing pressure. It was also observed that the coupling decreased for increasing power. We conclude that this must be caused by a change in the screening length, due to an increase in plasma density, or due to an increase in the dust thermal motion, since the dust charge actually increases with power.

The shape of the electron density profile in the experiment is well represented by the model, but the ion density profile measured in the modified GEC cell is very different from the measured electron density profiles, having its maximum in the center and falling off towards the electrode edge, rather than having an off-axis maximum. This results in a large positive space charge in the center of the discharge, while the space charge is more negative towards the edge. The model also predicts higher ion densities, resulting in an electropositive space charge, but the difference between electron and ion density is much smaller in the model than found in the experiment. A similar study in the original GEC cell \cite{GEC3} showed that the electron density and the ion density both have the same off-axis maximum, which is distinctly different from our finding. It did conclude that the ion density was higher than the electron density, so that the original GEC cell was also found to be electropositive at similar settings.

The plasma potential profiles obtained in the model are close to the measured profiles. The difference lies in the slope of the profiles. For higher pressures, the measured profiles result in electric fields pointing radially outwards, while lower pressures result in profiles pointing inwards. The model only gives profiles resulting in inward pointing electric fields, because for all pressures the net space charge has a clear positive maximum value towards the electrode edge. 

For dust particles immersed in the discharge, the observed change in the net radial force for different powers would mean an increase in radial compression for higher pressures. It is quite likely that the change in net force (even though the net forces are small) could be seen as a decrease in the inter-particle distance at higher pressures, due to the increased radial compression. Even though it is not directly obvious that the measured profiles remain the same towards the powered electrode below, it is likely that the radial confining electric field induced by the cutout in the electrode is affected by the change in the plasma radial electric field, due to a change in gas pressure. Therefore, the shape of a suspended dust cloud and the amount of dust particles that can be confined at a given height above the radial cutout might very well change dramatically for different neutral gas pressures. 

In short, the biggest discrepancy between the model and the experimental results are the plasma density profiles. It is likely that this discrepancy has the same basis as the discrepancy between the absorbed and the dissipated power, and has to do with the behaviour of the electron-neutral momentum transfer and its behavior with pressure. Overall, the local parameters seem to be close to the local parameters determined in the original GEC cell. Of special note, the change in the radial dust charge profile for a change in pressure, and the different shape of the ion density profile might be due to the change in geometry with respect to the original GEC cell. This means that despite the excellent agreement between the modified cell and the original cell when global parameters are considered, differences in the local profiles are still to be expected.

\subsection{Future work.}

The linear dependence of the model mobilities on neutral gas pressure is correct for low pressures, $P_{gas} \leq$ 10 mTorr, but a transition occurs for slightly higher pressures \cite{Godyak}, which means that an updated form for the electron-/ion-mobility is required. Using the obtained power dissipation data to obtain a better fit for the mobilities for a wider range of pressures is a possibility.

Obtaining two- or three-dimensional profiles of the plasma parameters, especially above the cutout on the lower electrode, is a logical next step in order to further benchmark our model, as well as to better understand the force balances for particles suspended in the parabolic confinement potential, especially for varying pressures and powers. These measurements could be made with a Zyvex S-100 nanomanipulator currently in use on another plasma cell in the CASPER laboratory, which can be equipped with different types of probes and can be very accurately steered throughout a large volume of the discharge \cite{Zyvex}.

The results from the discussed fluid model can be used as input and/or boundary conditions for MD models of collections of dust particles, that take the interaction potential between the charged particles into account. In this manner, dust crystal properties, such as the coupling parameter and bond angle, could be computed not only for random input parameters, but for plasma parameters consistent with observations. Phase transitions observed in MD simulations and experiments could thus be linked to the plasma parameters through the use of the fluid model, a good example of which is shown in \cite{Basner2009}.

\section{Summary}

We used a 2D fluid model to simulate a modified GEC reference cell used for dusty plasma experiments. We obtained the global characteristics in the form of the dissipated power and the natural DC bias. The model geometry is in excellent agreement with the modified geometry, showing that only a small fraction of the grounded surface area interacts with the plasma. The absorbed power versus root-mean-square potential behavior is well reproduced, but a difference in the behavior versus pressure indicates that an updated form for the charged particle mobilities is necessary. Such an update could be reproduced from a fit of the absorbed power for different pressures, which should give the plasma impedance.

The local dust charge, plasma densities, plasma potential and radial force profiles were obtained from the data and the model. The measured dust charge was higher than the dust charge obtained with the model, by a factor of 2. The dust charge profile shows a distinct dip and peak near the edge of the cutout for higher pressures. The ion density profile is quite different from the electron density profile, with a maximum value in the center and falling off towards the outer edge. In the original GEC cell, the ion density profiles were reported to be similar in shape to the electron density profile. The plasma potential profile at the midplane changes such that a radially confining electric field is produced at higher pressures. This might very well change the confining electric field closer to the powered electrode as well, resulting in a significant change in the radial confinement and the dust crystals formed in the sheath above the cutout.

Overall, our study shows that similar behaviour of global parameters, usually easily measured from externally accessible electronic signals, does not necessarily guarantee similarity of the local parameters, which are usually only accessible by measurements in-situ. This means that different modifications to the GEC reference cell destroy the standardization, meaning that for every modified GEC cell the pressure-power behavior has to be catalogued once again.

\end{document}